\documentclass[apj]{emulateapj}
\usepackage{amsmath}
\usepackage{mathrsfs}

\newcommand{\msun}{$M_{\odot}$}
\newcommand{\eg}{{\it e.g.}}
\newcommand{\ie}{{\it i.e.}}
\newcommand{\cmt}{cm$^{-3}$}

\begin{document}
\slugcomment{Accepted for publication in the Astrophysical Journal}
\shorttitle{On the Evolution of the DCMF}
\shortauthors{Swift \& Williams}
\title{On the Evolution of the Dense Core Mass Function}
\author{Jonathan J. Swift \& Jonathan P. Williams}
\email{js@ifa.hawaii.edu, jpw@ifa.hawaii.edu}
\affil{Institute for Astronomy, 2680 Woodlawn Dr., Honolulu, HI
96822-1897 }

\begin{abstract}
The mass distributions of dense cores in star-forming regions are
measured to have a shape similar to the initial mass function of
stars. This has been generally interpreted to mean that the
constituent cores will form individual stars or stellar systems at a
nearly constant star formation efficiency. This article presents a
series of numerical experiments evolving distributions of dense cores
into stars to quantify the effects of stellar multiplicity, global
core fragmentation, and a varying star formation efficiency. We find
that the different evolutionary schemes have an overall small effect
on the shape of the resultant distribution of stars. Our results
imply that at the current level of observational accuracy the
comparison between the mass functions of dense cores and stars alone
is insufficient to discern between different evolutionary
models. Observations over a wide range of mass scales including the
high or low-mass tails of these distributions have the largest
potential for discerning between different core evolutionary schemes.
\end{abstract}

\keywords{stars: formation --- ISM: clouds ---  ISM: structure } 

\section{Motivation} \label{sMot}
The mass of a star is the single most important parameter in
determining how it will interact with its environment, how long it
will live, and the nature of its death. Therefore the distribution of
masses of newly formed stars---the initial mass function~(IMF)---has
far reaching implications for the evolution of the cosmos.
The IMF is usually assumed to be universal, with a shape described by
a power-law above $\sim 1$\,\msun\ \citep[see][]{sca86,sca05,kro02}, and a
log-normal below \citep[see][]{cha03}.

Stars form from molecular clouds, and therefore their nature and
characteristics define the initial conditions for star formation. The
hierarchical density and velocity structure of molecular clouds is
indicative of supersonic turbulence
\citep{lar81,fal90,wil00}. However, cold, dense regions of quiescent
gas, or ``cores,'' found within Galactic clouds are believed to be the
sites of future low-mass star formation \citep{mye83a,ben89,ladd91}. 

A large 1.3\,mm continuum survey of dense cores in the 
$\rho$\,Ophiuchus star-forming region by \cite{mot98}
revealed a core mass distribution with a declining power-law slope
similar to the IMF. This led the authors to conclude that
the cores observed in thermal dust emission are the direct
progenitors of individual stars or stellar systems. The dense core
mass function (DCMF) in other regions has also exhibited similarities 
to the IMF \citep[\eg,][]{tes98,joh01,rei06,eno07,nut07}. Cores in the
Pipe Nebula identified through dust extinction show a 
turnover at masses about a factor of 3 higher than the IMF of the
Trapezium cluster leading to the interpretation that
cores evolve with a nearly constant star formation efficiency of $\sim
30$\% \citep{alv07}.

As first suggested by \cite{vaz94}, the density probability
distribution function (PDF) in isothermal turbulent flows is expected
to be log-normal \citep[see][for a review]{elm04} providing
theoretical means to produce the low-mass end of the DCMF. The
power-law tail of the DCMF can also be explained in terms of
post-shock gas within a turbulent medium \citep{pad02}, or by
deviations from isothermality \citep{sca_etal98}. In addition, the
star formation efficiency of cores is theoretically expected to be
nearly constant and lie between 30\%--50\% if outflows from protostars
are the primary mediating factor \citep{mat00}. 

These results support a one-to-one or nearly one-to-one relationship
between dense cores and the future stars to form from them. However,
the similarity of the DCMF to the IMF remains the only piece of
{\em observational} evidence for this kind of relationship. Meanwhile,
there are several reasons to think that this one-to-one relationship
may not hold.

It is clear that some dense cores must fragment to produce the large
fraction of observed multiple stellar systems \citep{duq91,goo07}, and
it is possible that several fragments per core may be necessary to
explain close binary systems \citep{ste03,goo05}. 
Unfortunately, most observations of pre-stellar cores are limited to
spatial resolutions orders of magnitude greater than
characteristic binary separations.

It is also difficult to determine what fraction of cores identified in
survey data will evolve into stars. In a recent study of cores in the
Pipe Nebula, \cite{lad07} found that the majority of cores are
gravitationally unbound, with only the highest mass cores appearing
destined to form stars. The most massive core in the Pipe Nebula,
Barnard 59, is the sole active core in the nebula and harbors an
association of $\sim 20$ young stars \citep{bro07}. 

Despite these open questions, the intriguing similarity between the
DCMF and the IMF remains. Given the numerous ways in which a core
could possibly evolve into a young star or stars, we construct
numerical simulations in an attempt to quantify the effects of
different core evolution scenarios on the resultant stellar IMF. We
follow this introduction with an outline of our methods and results in
\S~\ref{sSim}. Section~\ref{sDisc} discusses difficulties in comparing
astronomical datasets and expands upon our results in consideration of
observational constraints. We then conclude in \S~\ref{sConc} and
present a brief look toward future studies. 

\section{Simulated Evolution of the DCMF} \label{sSim}
\subsection{The Models} \label{sModels}
There are several mechanisms by which dense cores may form from more
diffuse molecular gas including ambipolar diffusion \citep{mou91},
thermal fragmentation \citep{jea61,lar85}, turbulent fragmentation
\citep{pad97}, or through triggering events \citep[see,
{\eg},][]{elm98}. \cite{pad02} derive a functional form for the PDF
of dense cores created by turbulent fragmentation that is in good
agreement with observations. 

All our simulations begin with a DCMF generated
from the PDF of their Equation 24. The role of turbulence in creating
dense cores is still under question \citep[see,
{\eg},][]{kir07}. However, the results of our experiments do not
depend critically on the validity of turbulent fragmentation 
since it is the difference between the DCMF and the IMF that we
are testing here, not the correctness of any particular formulation of 
the DCMF. We can thus take advantage of the convenient
analytical form for the DCMF provided by this formalism.

The simulations begin with a DCMF having a log-normal peak at
$\mu_c = 1.3$\,\msun, a dispersion $\sigma_c = 0.37$\,dex, and a
power-law tail at masses greater than a few \msun\  
with an index $x_c = 1.3$ (${\rm dN} \propto m^{-x_c}\,{\rm
d}\log{m}$), nearly equal to the Salpeter slope of the IMF
\citep[1.35;][]{sal55}. This shape is achieved using an Alfv\'{e}n
Mach number, $M_A = 5$, a kinetic temperature, $T_K = 10$\,K, an
average particle density, $n_0 = 1000$\,\cmt, and $\beta = 1.7$, where
$E_k \propto k^{-\beta}$ is the turbulent energy spectrum.

The models evolve cores with masses from $\log (M/M_\odot) = -2.5$ to
3.0 in increments of $\Delta \log M = 0.1$\,dex according to the
prescriptions described below. Each core mass bin maps to a stellar
probability distribution created numerically through 1000
repetitions. The stellar probability distributions generated from 
each core mass bin multiplied by the corresponding value of the core
PDF sum to create the final IMF. The characterizing parameters of the
final stellar distributions are $\mu_s$, $\sigma_s$, and $x_s$,
analogous to the DCMF.

\vspace{0.25cm}
\noindent
{\it One-to-One:} This simple model serves as a
comparison model for other evolutionary schemes and is labeled
{\sc ref}. Each core forms a single star with constant star formation
efficiency ${\rm SFE} = 0.3$. This creates an IMF with the precise
shape of the DCMF but shifted in $\log M$ by $\sim -0.5$\,dex.

\vspace{0.25cm}
\noindent
{\it Variable Star Formation Efficiency:} This model called {\sc
sfevar} evolves each core into a single star with uniformly
random ${\rm SFE} \in [0,1]$.  

\vspace{0.25cm}
\noindent
{\it Multiplicity:} These models explore two different multiplicity
scenarios motivated by observations. For model {\sc mult1}, each core
is converted into a single, binary, triple or quadruple system with a
probability of 57\%, 37\%, 4\% and 1\% respectively to produce stellar
systems with multiplicity in accord with solar-type field stars
\citep{duq91}. The system components are assigned uniformly random
mass ratios and then scaled such that the total mass in the system
equals the total core mass multiplied by a constant ${\rm SFE} =
0.3$. 

A mass dependency is embedded in model {\sc mult2} where only binaries
are considered, and the probability for binarity increases linearly
with mass from 10\% for core masses $\le 0.03$\,\msun\ to 100\% for
core masses $\ge 100$\,\msun. The mass ratio between pairs is
uniformly random.

\vspace{0.25cm}
\noindent
{\it Fragmentation:} Two different fragmentation schemes are explored
by these models. For model {\sc fragpdf}, each core is fragmented into
smaller cores repeatedly with masses drawn from the initial core mass
PDF until no mass remains in the original core. This model mimics
further turbulent fragmentation of observed cores. Model {\sc fraguni}
follows this same prescription, but draws the masses of fragments
randomly with uniform probability between 0 and the mass of the
original core. 

\vspace{0.25cm}
\noindent
{\it Composite:} The composite model, {\sc comp}, combines the
evolutionary formulas of all the above models. It applies a random
star formation efficiency, a {\sc mult1} multiplicity scheme, and
fragmentation according to model {\sc fragpdf}.  

\subsection{Results} \label{sRes}
Table~\ref{table} reports the results from all our models. 
All stellar distributions begin with a DCMF having $x_c = 1.3$,
$\sigma_c = 0.37$\,dex, and $\mu_c = 1.3$\,\msun. The slopes of the
stellar distributions are derived from power-law fits at masses
greater than $\mu_s + 2\sigma_s$, where the peaks and widths are
measured directly.
\begin{deluxetable}{lccc} 
  \tablecolumns{4} 
  \tablewidth{2.7in}
  \tablecaption{Model Results \label{table}}
    \tablehead{
      \colhead{Model} & 
      \colhead{$x_s$} &
      \colhead{$\sigma_s$} &    
      \colhead{$\mu_s$} \\
      \colhead{} &
      \colhead{} &
      \colhead{(dex)} &
      \colhead{(\msun)} } 
  \startdata 
  \multicolumn{4}{c}{Control Model} \\
  \cline{1-4} \vspace{-5pt}\\
  {\sc ref}\dotfill  & 1.3 & 0.37 & 0.39 \\
  \cutinhead{Variable SFE Model}
  {\sc sfevar}\dotfill  & 1.3 & 0.48 & 0.63 \\
  \cutinhead{Multiplicity Models}
  {\sc mult1}\dotfill  & 1.3 & 0.46 & 0.29 \\
  {\sc mult2}\dotfill  & 1.3 & 0.40 & 0.42 \\
  \cutinhead{Fragmentation Models}
  {\sc fragpdf}\dotfill  & 2.0 & 0.34 & 0.30 \\
  {\sc fraguni}\dotfill  & 1.3 & 0.51 & 0.23 \\ 
  \cutinhead{Composite Model}
  {\sc comp}\dotfill  & 2.0 & 0.53 & 0.34
  \enddata 
\end{deluxetable}

A variable star formation efficiency does not change the power-law slope
of the resulting stellar distribution from the initial DCMF. The peak
mass of the stellar distribution, $\mu_s$ is lower than $\mu_c$ by the
expected factor of 2, and the spread in SFE broadens the distribution
by 0.11\,dex. 

\begin{figure}
\centering
\includegraphics[angle=90,width=3.4in]{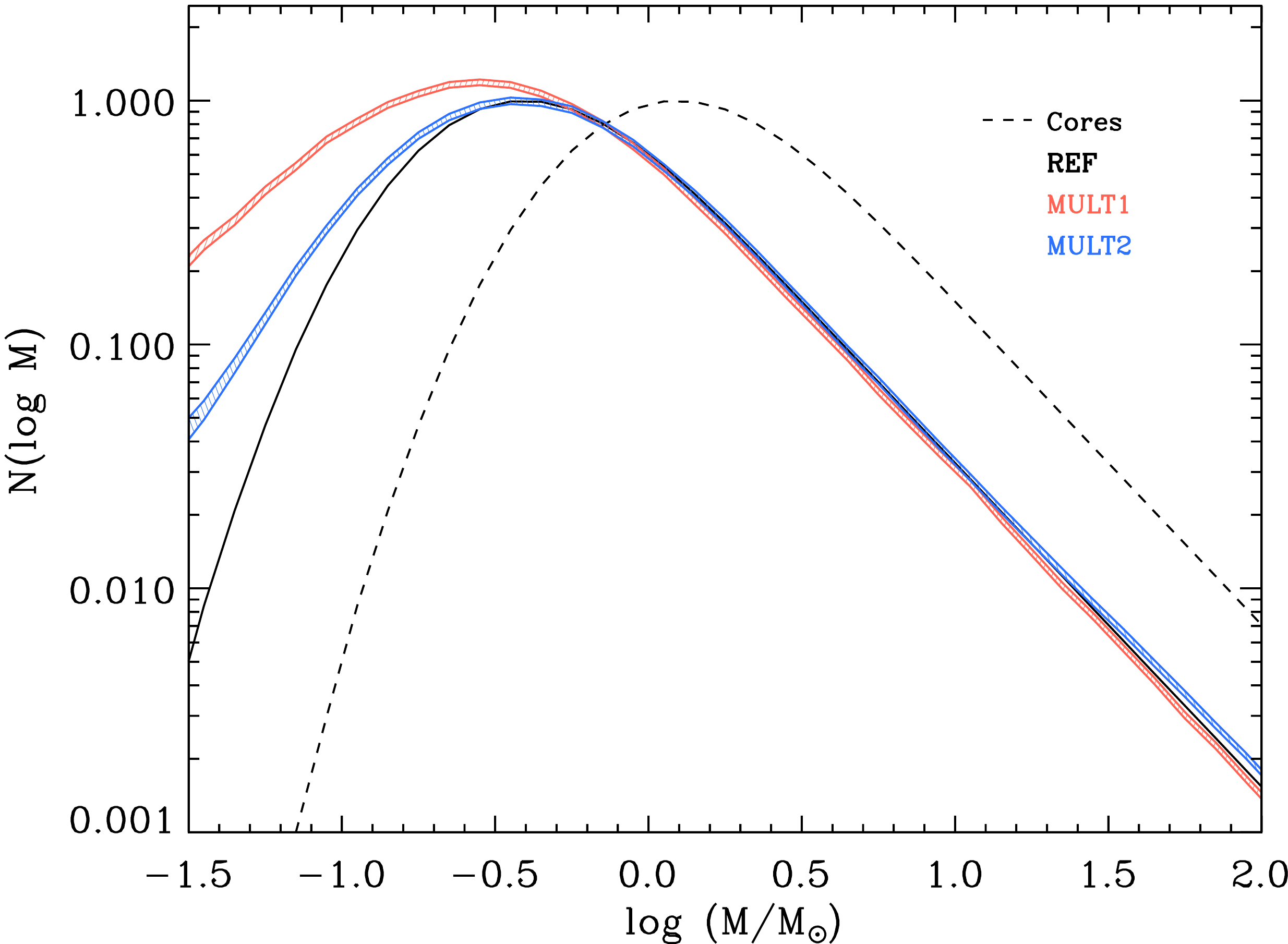}
\caption{Distributions of original core masses ({\it black dashed
    line}) and resultant stellar masses for a simple one-to-one core
  to star relationship ({\it black}), for a stellar distribution with
  a multiplicity in accord with \cite{duq91} ({\it red}), and a binary
  probability increasing with core mass ({\it blue}). 
  \label{multiplicity}} 
\end{figure}

Figure~\ref{multiplicity} shows the results of multiplicity models
{\sc mult1} and {\sc mult2} in red and blue, respectively. Model {\sc
ref} is shown for reference. The effect of stellar multiplicity has  
a negligible effect on the power-law portion of the IMF for either
model. The creation of low-mass companions in model {\sc mult1}
broadens the peak of the IMF by 0.09\,dex and also shifts $\mu_s$
toward lower mass by $\sim 25$\% in comparison to model {\sc ref}.  

The redistribution of mass takes place mostly in the self-similar part
of the DCMF for model {\sc mult2}. Therefore there is little change in
the overall shape of the resultant IMF. The mean binary fraction for
all stars created in model {\sc mult2} is $\sim 30$\%.

\begin{figure}[!hb]
\centering
\includegraphics[angle=90,width=3.4in]{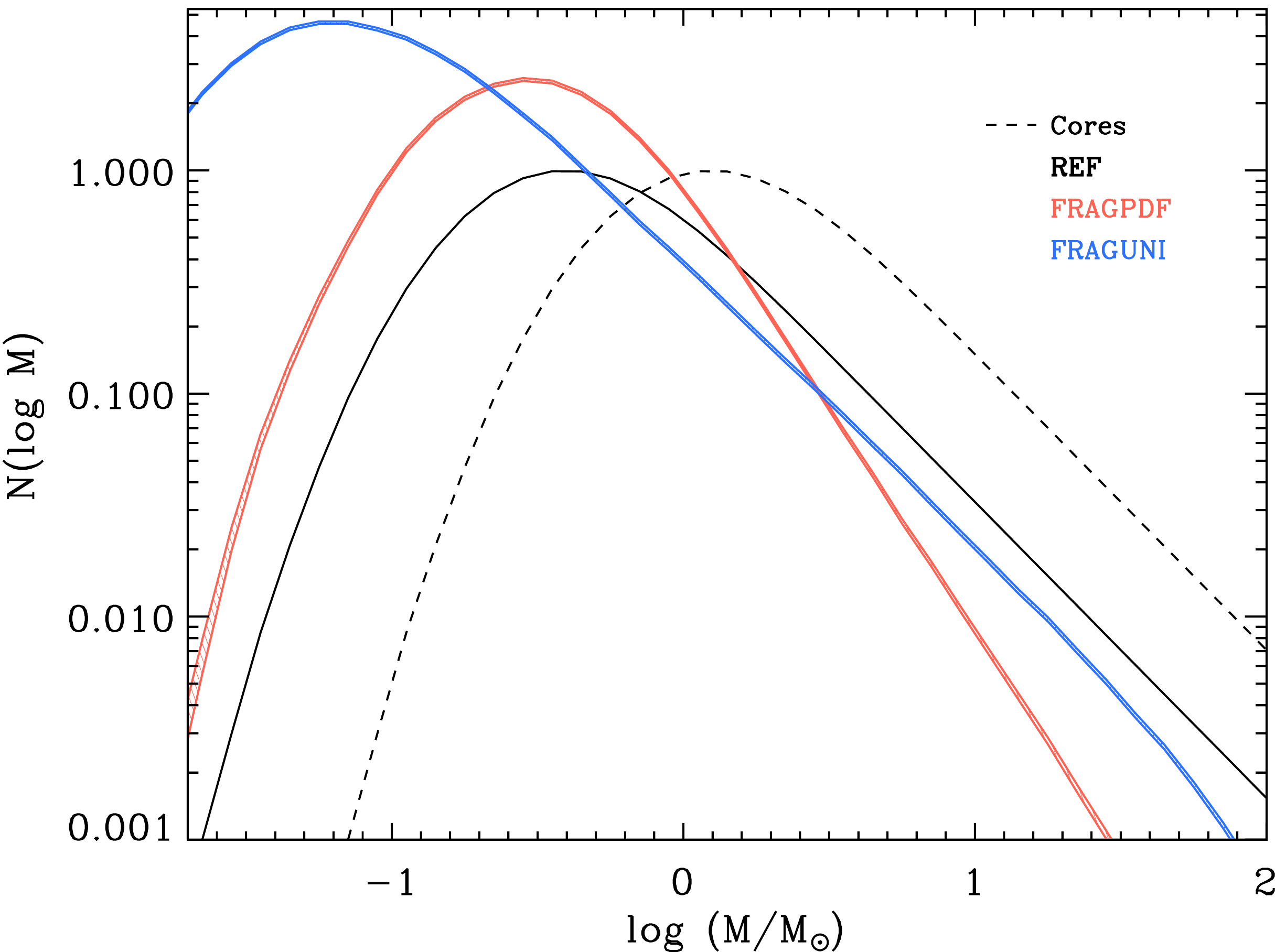}
\caption{Mass distributions from our fragmentation
  models. The {\it black dashed} and {\it solid} curves are the same
  as for Figure~\ref{multiplicity}. The {\it red} distribution
  describes stellar masses resulting from core fragments with masses
  drawn from the original core PDF, while the {\it blue} distribution
  represents uniformly random fragment masses.  \label{fragmentation}} 
\end{figure}

The results of our fragmentation models are shown graphically in
Figure~\ref{fragmentation}. Again, model {\sc ref} is  
shown for reference. The cores of model {\sc fragpdf}
fragment into smaller cores with masses preferentially near
$\mu_c$. Therefore the higher mass cores have a higher number
of fragments---up to $\sim 100$ for the most massive cores. This
creates a narrower stellar distribution with a significantly higher
peak compared to model {\sc ref} and a steeper power-law slope by
$0.7$ in the index.  

The stellar distribution of model {\sc fraguni} has a power-law tail
identical to the DCMF. The downward turn of the blue curve at the
high-mass end in Figure~\ref{fragmentation} is due to the limited mass
range over which we applied this fragmentation scheme. The randomly
assigned fragment masses do not change the self-similar part of the
stellar distribution, but the excess of low-mass stars generated by
cores spanning the entire DCMF widens the resultant IMF by
0.14\,dex. The mean number of fragments per core in this scheme is
$\sim 2$. 

The composite model, {\sc comp}, combines stellar multiplicity, core
fragmentation, and a random star formation efficiency. 
While the preferred mass of the fragmentation creates a steep
power-law tail as in model {\sc fragpdf}, the broader peak due to
multiplicity and a variable SFE create a shape consistent with the
original core PDF over a wider mass range than {\sc fragpdf} (see
Figure~\ref{fragcomp} below). 

\vspace{0.25cm}
\section{Discussion} \label{sDisc}
\subsection{Observational Difficulties}
The different core evolution schemes produce stellar distributions
that are clearly discernible in our theoretical modeling. But these
simulated data offer the luxury of complete control. When comparing  
real, astronomical datasets of cores and stars, several difficulties
arise.

\subsubsection{Obtaining a Representative Sample of Cores}
The unbiased identification of an ensemble of dense
cores within molecular clouds that will definitively form stars is a
difficult task. Many automated core or clump finding algorithms exist
that produce reasonable results from dust continuum, dust extinction or
molecular line data \citep[{\eg}][]{stu90,will94}. However, 
these methods offer no measure of systematic errors in identifying
bona fide pre-stellar cores which may dominate the Poisson errors
assumed in analyses. 

Once a core forms a protostar, it is not clear how the mass of the
remaining core is relevant to the IMF. Cores with embedded stars can
therefore be excluded from an ensemble with the use of sensitive
infrared observations, {\eg}, using {\it Spitzer} \citep{eva03}. The
remaining starless cores, however, may not all form stars in the
future.

Recent studies of the Perseus star-forming region show that starless
cores tend to be less massive than cores with stars (although there
exist many low-mass cores that harbor embedded sources)
\citep[see][]{jor07,hat07}. This could mean that many of the low-mass
cores included in their samples are transient structures or perhaps
still accreting material. Molecular line data can supplement dust maps
to determine the gravitational boundedness of cores, and hence the
likelihood of them eventually forming stars. In a study of the Pipe
nebula cores, \cite{lad07} find that only the $\sim 25$\% most massive
cores are gravitationally bound, and \cite{joh00} find that a
majority of cores in Ophiuchus are stable against gravity. However,
molecular line studies of NGC\,1333 and Ophiuchus find that cores in
those regions extending down to masses of $\sim 0.1$\,\msun\ are
likely to be bound \citep{wal07,and07}.

Low-mass, starless cores that appear gravitationally unbound might
also be bound by external pressure, and are therefore potentially
stellar precursors. In this case, the possibility of different
evolutionary timescales for different cores in the sample may need to
be considered \citep[see][for details]{cla07}.

\subsubsection{Comparing DCMFs to the IMF}
It is impossible to measure a DCMF for an ensemble of cores as well as
an IMF for the stars that formed from them. Variations in the DCMF
({\eg}, compare Motte et al. 1998 \and Nutter \& Ward-Thompson 2007)
and the IMF \citep{cha03} from region to region therefore must be
considered when comparing these distributions. Moreover, the total
number of stars to form from a distribution of dense cores can never
be measured directly. This means that the height of, or total area
under, the stellar IMF cannot be used to discern between core
evolution models.  

Observations of cores in star forming regions are currently limited to
samples of less than a few hundred
\citep[{\eg},][]{eno07,nut07,alv07}. Using our core PDF, we generate
simulated data for a sample of 300 cores. This number was chosen to
give a number of stars per $\Delta\log M = 0.2$\,dex bin near the peak
of the DCMF comparable to modern-day observations.  

\begin{figure}
\centering
\includegraphics[angle=90,width=3.4in]{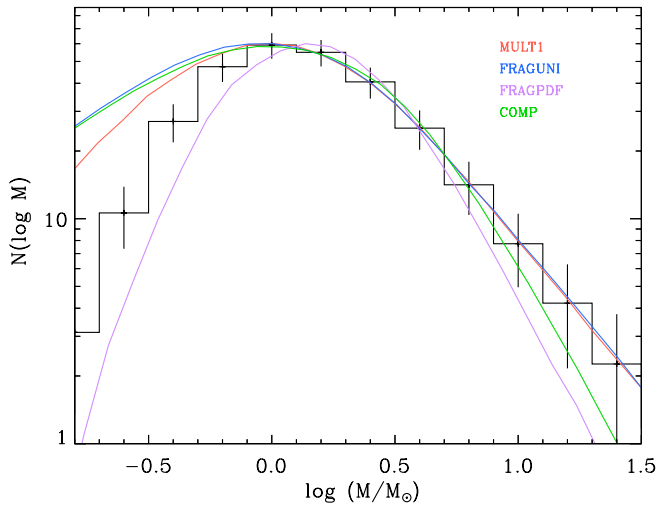}
\caption{Comparison between simulated observations of the DCMF and the
  synthetic IMFs of several of our models. The error bars 
  represent $\sqrt{N}$ measurement errors.  \label{fragcomp}}
\end{figure}

Figure~\ref{fragcomp} shows the comparison between these simulated
observations and numerous results from our models. The errors on the
simulated data correspond to $\sqrt{N}$ from counting statistics.
The synthetic IMFs are rescaled by a multiplicative factor and shifted
by an additive factor in $\log M$ that produces the lowest $\chi^2$
value between $-0.1$ and $1.1$ in $\log (M/M_\sun)$. All synthetic
IMFs produce $\chi^2$ values less than one except model {\sc fragpdf}
for which $\chi^2 = 3.6$. 

Our models produce the largest variation in the parameters $\mu_s$ and
$\sigma_s$, while only mass dependent core evolution affects $x_s$.
Fragmentation affects the width of the IMF and tends to 
lower the characteristic mass of the IMF compared to a simple
one-to-one core evolution scenario. Variation in the SFE broadens the
IMF while the mean SFE for an ensemble of cores affects the
characteristic mass of the IMF. 
Additionally, mass dependency may exist in either core fragmentation
or in the SFE of cores. Therefore there is significant degeneracy
between the nature of fragmentation and SFE in determining the shape
of the IMF.   

To break this degeneracy, accurate measurements of DCMFs and the IMF
over $\gtrsim 2$ orders of magnitude in mass must be made, where the
most leverage is achieved at the high and low-mass ends of these
distributions. This data must then be supplemented with independent
observations, such as comparisons between the total mass in cores and
total mass in stars to constrain the SFE, or spatial correlation
functions of stars and cores to constrain the fragmentation. 

\subsection{Observational Constraints}
The SFE of dense cores in Perseus has been recently determined to be 
between 10\%--15\% \citep{jor07}. This result implies
that the horizontal shift between measured DCMFs and the stellar
IMF should be a factor of order 10. However, our models show that it
is not the SFE alone that contributes to the horizontal shift between
DCMFs and the IMF; fragmentation can also contribute.

The SFE in model {\sc mult1} derived by direct comparison of the DCMF
and IMF appears $\sim 27$\% lower than the true SFE of $0.3$ due
merely to the minimal fragmentation required to reproduce the observed
multiplicity of field stars in the Galaxy. For model {\sc fraguni} the
apparent SFE drops to $0.17$. Fragmentation effects could therefore
lead to an underestimation of the SFE by as much as 40\% for a given
measured DCMF. Alternatively, it is shown through model {\sc mult2}
that the opposite systematic effect could be an issue if higher mass
stars preferentially form multiple systems; {\ie}, an overestimation
of the inferred SFE from a measured DCMF.

Without considering the possibility of core fragmentation, the
horizontal shifts between measured DCMFs and the IMF imply SFEs of
$\gtrsim 50$\% in Ophiuchus \citep{mot98,and07}, to $\sim 30$\% in the
Pipe Nebula \citep{alv07}, to $\sim 6$\% in Orion \citep{nut07}. This 
wide spread may be partially due to a varying degrees of core
fragmentation in the different regions. 
However, it is not clear to what degree this discrepancy is due to
observational biases. For instance, the poorer physical resolution
of observations toward distant regions may artificially drive the DCMF
peak to higher masses because of confusion, thus lowering the inferred
SFE.

The number of cores containing multiple embedded protostars within a
star-forming region is related to the amount of core fragmentation
that has taken place there. From the data of \citet[][Table~1]{jor07},
anywhere from 4\% to 23\% of cores have formed multiple protostars
with separations between $\sim 1500$\,AU and $\sim 4000$\,AU. 
This is a lower limit to the amount of core fragmentation in
Perseus primarily because $\sim 40$\% of cores are likely to fragment on
scales smaller than 1500\,AU to produce a multiplicity in accord
observations of field stars \citep{duq91}.

However, for comparison, we calculate from our fragmentation models an
upper limit to the number of Perseus cores expected to be observed
with multiple protostars. Thirty percent of cores with masses between
0.1 and 4.9\,\msun\ \citep{kir06} fragment into multiple cores
according to the {\sc fragpdf} model, while 42\% of the cores from the
{\sc fraguni} model potentially produce cores with multiple protostars. 

\section{Conclusions} \label{sConc}
If the structure and conditions within molecular clouds indeed
determine the stellar IMF, the similarity between measured DCMFs and
the IMF is likely a reflection of this relationship. It is clear from
past studies that some of the cores identified in star-forming regions
are the direct progenitors of stars or stellar systems. However, our
simulations have shown that the overall shape of the IMF is robust
against different core evolution scenarios for core masses between
$\sim 0.5$ and $10$\,\msun. In light of the uncertainties in identifying
a representative sample of cores in astronomical data, a direct
comparison between DCMFs in this mass range and the IMF cannot alone
imply any particular evolutionary path from an ensemble of cores to a
future generation of stars.

The peak mass in our simulated IMFs, $\mu_s$, is affected by both the
SFE and the fragmentation scheme adopted. Estimations of the SFE by
comparing the turnover in the DCMF to the IMF could be off by as much
as $\sim 40$\% without considering the possibility of core
fragmentation (including multiplicity). 
The widths of our simulated IMFs, $\sigma_s$, are also affected both
by variation in the SFE and fragmentation. Therefore, independent
constraints on the SFE ({\eg}, through comparisons between total core
mass and total stellar mass) or fragmentation ({\eg}, using spatial
correlation functions of cores and stars) must be used to discern
between these two effects given accurate measurements of DCMFs and the
IMF over $\gtrsim 2$ orders of magnitude in mass.

Differences in the power-law tails of the DCMF and IMF are sensitive
to mass dependencies in core evolution. Using a fragmentation scheme
with a preferential mass scale we achieve a power-law slope change of
$0.7$, though other fragmentation scenarios produce no
noticeable difference.

\subsection{Future Studies} \label{sFuture}
The massive clouds identified through $8\,\mu$m extinction against the
Galactic background---infrared dark clouds
\citep[IRDCs;][]{per96,ega98}---offer a promising target list from
which to derive DCMFs over a wide mass range extending to higher
masses than have been previously measured. These sources, however, are
typically distant making individual cores difficult to detect and
resolve. ALMA may be needed to produce useful DCMFs from these
sources.   

To constrain the low-mass end of the DCMF, it is necessary to make
sensitive, high-resolution observations of the closest star-forming
regions. Current facilities have the resolution to detect individual
cores, and instruments such as SCUBA2 may extend considerably the DCMF
to lower masses over wide fields. Together with molecular line data
and complete surveys of late-type stars and brown dwarfs in a variety
of star-forming regions, the statistics at the low-mass ends of DCMFs
and IMFs may soon be sufficient to draw detailed conclusions regarding
the relationship between the DCMF and the IMF.  

\acknowledgments
We thank Doug Johnstone, the referee, for spurring a better version of
this work. J. S. would also like to thank N. Evans for various short,
but motivating discussions on this topic. J. P. W. acknowledges
support from NSF grant AST-0324328.


\begin{thebibliography}{47}
\expandafter\ifx\csname natexlab\endcsname\relax\def\natexlab#1{#1}\fi

\bibitem[{{Alves} {et~al.}(2007){Alves}, {Lombardi}, \& {Lada}}]{alv07}
{Alves}, J., {Lombardi}, M., \& {Lada}, C.~J. 2007, \aap, 462, L17

\bibitem[{{Andr{\'e}} {et~al.}(2007){Andr{\'e}}, {Belloche}, {Motte}, \&
  {Peretto}}]{and07}
{Andr{\'e}}, P., {Belloche}, A., {Motte}, F., \& {Peretto}, N. 2007, \aap, 472,
  519

\bibitem[{{Benson} \& {Myers}(1989)}]{ben89}
{Benson}, P.~J. \& {Myers}, P.~C. 1989, \apjs, 71, 89

\bibitem[{{Brooke} {et~al.}(2007){Brooke}, {Huard}, {Bourke}, {Boogert},
  {Allen}, {Blake}, {Evans}, {Harvey}, {Koerner}, {Mundy}, {Myers}, {Padgett},
  {Sargent}, {Stapelfeldt}, {van Dishoeck}, {Chapman}, {Cieza}, {Dunham},
  {Lai}, {Porras}, {Spiesman}, {Teuben}, {Young}, {Wahhaj}, \& {Lee}}]{bro07}
{Brooke}, T.~Y., {Huard}, T.~L., {Bourke}, T.~L., {Boogert}, A.~C.~A., {Allen},
  L.~E., {Blake}, G.~A., {Evans}, II, N.~J., {Harvey}, P.~M., {Koerner}, D.~W.,
  {Mundy}, L.~G., {Myers}, P.~C., {Padgett}, D.~L., {Sargent}, A.~I.,
  {Stapelfeldt}, K.~R., {van Dishoeck}, E.~F., {Chapman}, N., {Cieza}, L.,
  {Dunham}, M.~M., {Lai}, S.-P., {Porras}, A., {Spiesman}, W., {Teuben}, P.~J.,
  {Young}, C.~H., {Wahhaj}, Z., \& {Lee}, C.~W. 2007, \apj, 655, 364

\bibitem[{{Chabrier}(2003)}]{cha03}
{Chabrier}, G. 2003, \pasp, 115, 763

\bibitem[{{Clark} {et~al.}(2007){Clark}, {Klessen}, \& {Bonnell}}]{cla07}
{Clark}, P.~C., {Klessen}, R.~S., \& {Bonnell}, I.~A. 2007, \mnras, 379, 57

\bibitem[{{Duquennoy} \& {Mayor}(1991)}]{duq91}
{Duquennoy}, A. \& {Mayor}, M. 1991, \aap, 248, 485

\bibitem[{{Egan} {et~al.}(1998){Egan}, {Shipman}, {Price}, {Carey}, {Clark}, \&
  {Cohen}}]{ega98}
{Egan}, M.~P., {Shipman}, R.~F., {Price}, S.~D., {Carey}, S.~J., {Clark},
  F.~O., \& {Cohen}, M. 1998, \apjl, 494, L199+

\bibitem[{{Elmegreen}(1998)}]{elm98}
{Elmegreen}, B.~G. 1998, in Astronomical Society of the Pacific Conference
  Series, Vol. 148, Origins, ed. C.~E. {Woodward}, J.~M. {Shull}, \& H.~A.
  {Thronson}, Jr., 150--+

\bibitem[{{Elmegreen} \& {Scalo}(2004)}]{elm04}
{Elmegreen}, B.~G. \& {Scalo}, J. 2004, \araa, 42, 211

\bibitem[{{Enoch} {et~al.}(2007){Enoch}, {Glenn}, {Evans}, {Sargent}, {Young},
  \& {Huard}}]{eno07}
{Enoch}, M.~L., {Glenn}, J., {Evans}, II, N.~J., {Sargent}, A.~I., {Young},
  K.~E., \& {Huard}, T.~L. 2007, \apj, 666, 982

\bibitem[{{Evans} {et~al.}(2003){Evans}, {Allen}, {Blake}, {Boogert}, {Bourke},
  {Harvey}, {Kessler}, {Koerner}, {Lee}, {Mundy}, {Myers}, {Padgett},
  {Pontoppidan}, {Sargent}, {Stapelfeldt}, {van Dishoeck}, {Young}, \&
  {Young}}]{eva03}
{Evans}, II, N.~J., {Allen}, L.~E., {Blake}, G.~A., {Boogert}, A.~C.~A.,
  {Bourke}, T., {Harvey}, P.~M., {Kessler}, J.~E., {Koerner}, D.~W., {Lee},
  C.~W., {Mundy}, L.~G., {Myers}, P.~C., {Padgett}, D.~L., {Pontoppidan}, K.,
  {Sargent}, A.~I., {Stapelfeldt}, K.~R., {van Dishoeck}, E.~F., {Young},
  C.~H., \& {Young}, K.~E. 2003, \pasp, 115, 965

\bibitem[{{Falgarone} \& {Phillips}(1990)}]{fal90}
{Falgarone}, E. \& {Phillips}, T.~G. 1990, \apj, 359, 344

\bibitem[{{Goodwin} \& {Kroupa}(2005)}]{goo05}
{Goodwin}, S.~P. \& {Kroupa}, P. 2005, \aap, 439, 565

\bibitem[{{Goodwin} {et~al.}(2007){Goodwin}, {Kroupa}, {Goodman}, \&
  {Burkert}}]{goo07}
{Goodwin}, S.~P., {Kroupa}, P., {Goodman}, A., \& {Burkert}, A. 2007, in
  Protostars and Planets V, ed. B.~{Reipurth}, D.~{Jewitt}, \& K.~{Keil},
  133--147

\bibitem[{{Hatchell} {et~al.}(2007){Hatchell}, {Fuller}, {Richer}, {Harries},
  \& {Ladd}}]{hat07}
{Hatchell}, J., {Fuller}, G.~A., {Richer}, J.~S., {Harries}, T.~J., \& {Ladd},
  E.~F. 2007, \aap, 468, 1009

\bibitem[{{Jeans}(1961)}]{jea61}
{Jeans}, J.~H. 1961, {Astronomy and cosmogony} (New York: Dover, 1961)

\bibitem[{{Johnstone} {et~al.}(2001){Johnstone}, {Fich}, {Mitchell}, \&
  {Moriarty-Schieven}}]{joh01}
{Johnstone}, D., {Fich}, M., {Mitchell}, G.~F., \& {Moriarty-Schieven}, G.
  2001, \apj, 559, 307

\bibitem[{{Johnstone} {et~al.}(2000){Johnstone}, {Wilson}, {Moriarty-Schieven},
  {Joncas}, {Smith}, {Gregersen}, \& {Fich}}]{joh00}
{Johnstone}, D., {Wilson}, C.~D., {Moriarty-Schieven}, G., {Joncas}, G.,
  {Smith}, G., {Gregersen}, E., \& {Fich}, M. 2000, \apj, 545, 327

\bibitem[{{J{\o}rgensen} {et~al.}(2007){J{\o}rgensen}, {Johnstone}, {Kirk}, \&
  {Myers}}]{jor07}
{J{\o}rgensen}, J.~K., {Johnstone}, D., {Kirk}, H., \& {Myers}, P.~C. 2007,
  \apj, 656, 293

\bibitem[{{Kirk} {et~al.}(2006){Kirk}, {Johnstone}, \& {Di Francesco}}]{kir06}
{Kirk}, H., {Johnstone}, D., \& {Di Francesco}, J. 2006, \apj, 646, 1009

\bibitem[{{Kirk} {et~al.}(2007){Kirk}, {Johnstone}, \& {Tafalla}}]{kir07}
{Kirk}, H., {Johnstone}, D., \& {Tafalla}, M. 2007, \apj, 668, 1042

\bibitem[{{Kroupa}(2002)}]{kro02}
{Kroupa}, P. 2002, Science, 295, 82

\bibitem[{{Lada} {et~al.}(2007){Lada}, {Muench}, {Rathborne}, {Alves}, \&
  {Lombardi}}]{lad07}
{Lada}, C.~J., {Muench}, A.~A., {Rathborne}, J.~M., {Alves}, J.~F., \&
  {Lombardi}, M. 2007, ArXiv e-prints, 709

\bibitem[{{Ladd} {et~al.}(1991){Ladd}, {Adams}, {Fuller}, {Myers}, {Casey},
  {Davidson}, {Harper}, \& {Padman}}]{ladd91}
{Ladd}, E.~F., {Adams}, F.~C., {Fuller}, G.~A., {Myers}, P.~C., {Casey}, S.,
  {Davidson}, J.~A., {Harper}, D.~A., \& {Padman}, R. 1991, \apj, 382, 555

\bibitem[{{Larson}(1981)}]{lar81}
{Larson}, R.~B. 1981, \mnras, 194, 809

\bibitem[{{Larson}(1985)}]{lar85}
---. 1985, \mnras, 214, 379

\bibitem[{{Matzner} \& {McKee}(2000)}]{mat00}
{Matzner}, C.~D. \& {McKee}, C.~F. 2000, \apj, 545, 364

\bibitem[{{Motte} {et~al.}(1998){Motte}, {Andre}, \& {Neri}}]{mot98}
{Motte}, F., {Andre}, P., \& {Neri}, R. 1998, \aap, 336, 150

\bibitem[{{Mouschovias}(1991)}]{mou91}
{Mouschovias}, T.~C. 1991, \apj, 373, 169

\bibitem[{{Myers} {et~al.}(1983){Myers}, {Linke}, \& {Benson}}]{mye83a}
{Myers}, P.~C., {Linke}, R.~A., \& {Benson}, P.~J. 1983, \apj, 264, 517

\bibitem[{{Nutter} \& {Ward-Thompson}(2007)}]{nut07}
{Nutter}, D. \& {Ward-Thompson}, D. 2007, \mnras, 374, 1413

\bibitem[{{Padoan} {et~al.}(1997){Padoan}, {Jones}, \& {Nordlund}}]{pad97}
{Padoan}, P., {Jones}, B.~J.~T., \& {Nordlund}, A.~P. 1997, \apj, 474, 730

\bibitem[{{Padoan} \& {Nordlund}(2002)}]{pad02}
{Padoan}, P. \& {Nordlund}, {\AA}. 2002, \apj, 576, 870

\bibitem[{{Perault} {et~al.}(1996){Perault}, {Omont}, {Simon}, {Seguin},
  {Ojha}, {Blommaert}, {Felli}, {Gilmore}, {Guglielmo}, {Habing}, {Price},
  {Robin}, {de Batz}, {Cesarsky}, {Elbaz}, {Epchtein}, {Fouque}, {Guest},
  {Levine}, {Pollock}, {Prusti}, {Siebenmorgen}, {Testi}, \& {Tiphene}}]{per96}
{Perault}, M., {Omont}, A., {Simon}, G., {Seguin}, P., {Ojha}, D., {Blommaert},
  J., {Felli}, M., {Gilmore}, G., {Guglielmo}, F., {Habing}, H., {Price}, S.,
  {Robin}, A., {de Batz}, B., {Cesarsky}, C., {Elbaz}, D., {Epchtein}, N.,
  {Fouque}, P., {Guest}, S., {Levine}, D., {Pollock}, A., {Prusti}, T.,
  {Siebenmorgen}, R., {Testi}, L., \& {Tiphene}, D. 1996, \aap, 315, L165

\bibitem[{{Reid} \& {Wilson}(2006)}]{rei06}
{Reid}, M.~A. \& {Wilson}, C.~D. 2006, \apj, 650, 970

\bibitem[{{Salpeter}(1955)}]{sal55}
{Salpeter}, E.~E. 1955, \apj, 121, 161

\bibitem[{{Scalo}(2005)}]{sca05}
{Scalo}, J. 2005, in Astrophysics and Space Science Library, Vol. 327, The
  Initial Mass Function 50 Years Later, ed. E.~{Corbelli}, F.~{Palla}, \&
  H.~{Zinnecker}, 23--+

\bibitem[{{Scalo} {et~al.}(1998){Scalo}, {Vazquez-Semadeni}, {Chappell}, \&
  {Passot}}]{sca_etal98}
{Scalo}, J., {Vazquez-Semadeni}, E., {Chappell}, D., \& {Passot}, T. 1998,
  \apj, 504, 835

\bibitem[{{Scalo}(1986)}]{sca86}
{Scalo}, J.~M. 1986, Fundam.~Cosmic Phys., Vol.~11, Nos.~1 - 3, p.~1 - 278, 11,
  1

\bibitem[{{Sterzik} {et~al.}(2003){Sterzik}, {Durisen}, \& {Zinnecker}}]{ste03}
{Sterzik}, M.~F., {Durisen}, R.~H., \& {Zinnecker}, H. 2003, \aap, 411, 91

\bibitem[{{Stutzki} \& {Guesten}(1990)}]{stu90}
{Stutzki}, J. \& {Guesten}, R. 1990, \apj, 356, 513

\bibitem[{{Testi} \& {Sargent}(1998)}]{tes98}
{Testi}, L. \& {Sargent}, A.~I. 1998, \apjl, 508, L91

\bibitem[{{Vazquez-Semadeni}(1994)}]{vaz94}
{Vazquez-Semadeni}, E. 1994, \apj, 423, 681

\bibitem[{{Walsh} {et~al.}(2007){Walsh}, {Myers}, {Di Francesco}, {Mohanty},
  {Bourke}, {Gutermuth}, \& {Wilner}}]{wal07}
{Walsh}, A.~J., {Myers}, P.~C., {Di Francesco}, J., {Mohanty}, S., {Bourke},
  T.~L., {Gutermuth}, R., \& {Wilner}, D. 2007, \apj, 655, 958

\bibitem[{{Williams} {et~al.}(2000){Williams}, {Blitz}, \& {McKee}}]{wil00}
{Williams}, J.~P., {Blitz}, L., \& {McKee}, C.~F. 2000, Protostars and Planets
  IV, 97

\bibitem[{{Williams} {et~al.}(1994){Williams}, {de Geus}, \& {Blitz}}]{will94}
{Williams}, J.~P., {de Geus}, E.~J., \& {Blitz}, L. 1994, \apj, 428, 693

\end{thebibliography}
\end{document}